\documentclass[aps,prd,preprint,draft,showpacs,nofootinbib]{revtex4}
\usepackage{graphicx,epsf,amssymb}
\newcommand{\be}{\begin{equation}} 
\newcommand{\ee}{\end{equation}}

\newcommand{\bea}{\begin{eqnarray}} 
\newcommand{\eea}{\end{eqnarray}}

\def\dblhat#1{\hskip 6pt%
\widehat {\vphantom{\raise 1pt\hbox{$#1$}}%
\rlap{$\kern -6.5pt #1$}%
{\smash{\scriptstyle\widehat{\vphantom{\raise 0.5pt\hbox{$#1$}}}}}%
}\hskip 4pt}

\newif\ifdraft
\drafttrue
\newif\ifpreprint
\preprinttrue

\def\sect#1{section~{\ref{#1}}}

\def\NeqFour{{\cal N}=4}
\def\NeqOne{{\cal N}=1}

\def\spa#1.#2{\left\langle#1\,#2\right\rangle}
\def\spb#1.#2{\left[#1\,#2\right]}
\def\sand#1.#2.#3{%
\left\langle\smash{#1}{\vphantom1}^{-}\right|{#2}%
\left|\smash{#3}{\vphantom1}^{-}\right\rangle}
\def\sandp#1.#2.#3{%
\left\langle\smash{#1}{\vphantom1}^{-}\right|{#2}%
\left|\smash{#3}{\vphantom1}^{+}\right\rangle}
\def\sandpp#1.#2.#3{%
\left\langle\smash{#1}{\vphantom1}^{+}\right|{#2}%
\left|\smash{#3}{\vphantom1}^{+}\right\rangle}
\def\sandpm#1.#2.#3{%
\left\langle\smash{#1}{\vphantom1}^{+}\right|{#2}%
\left|\smash{#3}{\vphantom1}^{-}\right\rangle}
\def\sandmp#1.#2.#3{%
\left\langle\smash{#1}{\vphantom1}^{-}\right|{#2}%
\left|\smash{#3}{\vphantom1}^{+}\right\rangle}
\def\sandmm#1.#2.#3{%
\left\langle\smash{#1}{\vphantom1}^{-}\right|{#2}%
\left|\smash{#3}{\vphantom1}^{-}\right\rangle}
\def\spab#1.#2.#3{\sandmm#1.#2.#3}
\def\spba#1.#2.#3{\sandpp#1.#2.#3}
\def\spaa#1.#2.#3.#4{\sandmp#1.{#2#3}.#4}
\def\spbb#1.#2.#3.#4{\sandpm#1.{#2#3}.#4}
\def\spash#1.#2{\spa{\smash{#1}}.{\smash{#2}}}

\newbox\charbox
\newbox\slabox
\def\s#1{{      % Feynman slash
        \setbox\charbox=\hbox{$#1$}
        \setbox\slabox=\hbox{$/$}
        \dimen\charbox=\ht\slabox
        \advance\dimen\charbox by -\dp\slabox
        \advance\dimen\charbox by -\ht\charbox
        \advance\dimen\charbox by \dp\charbox
        \divide\dimen\charbox by 2
        \raise-\dimen\charbox\hbox to \wd\charbox{\hss/\hss}
        \llap{$#1$}
}}

\newbox\SlashedBox 
\def\slashed#1{\setbox\SlashedBox=\hbox{#1}
\hbox to 0pt{\hbox to 1\wd\SlashedBox{\hfil/\hfil}\hss}{#1}}
\def\hboxtosizeof#1#2{\setbox\SlashedBox=\hbox{#1}
\hbox to 1\wd\SlashedBox{#2}}
\def\mathslashed#1{\setbox\SlashedBox=\hbox{$#1$}
\hbox to 0pt{\hbox to 1\wd\SlashedBox{\hfil/\hfil}\hss}#1}
\def\mathslashedS#1#2{\setbox\SlashedBox=\hbox{${#1#2}$}
\hbox to 0pt{\hbox to 1\wd\SlashedBox{\hfil${#1/}$\hfil}\hss}{#1#2}}

\def\eqn#1{eq.~(\ref{#1})}

\def\eqns#1#2{eqs.~(\ref{#1}) and~(\ref{#2})}

\def\qb{{\overline {\kern-0.7pt q\kern -0.7pt}}}

\def\e{\epsilon}

\def\sign{{\mathop{\rm sign}\nolimits}}

\def\Ord{{\cal O}}

\def\sandp#1.#2.#3{%
\left\langle\smash{#1}{\vphantom1}^{+}\right|{#2}%
\left|\smash{#3}{\vphantom1}^{+}\right\rangle}
\def\ksl{\s{k}}
\def\Ksl{\s{K}}
\def\Ksls{\mathslashedS{\scriptscriptstyle}{K}}
\def\tlambda{{\tilde\lambda}}

\newbox\ourfigbox
\def\SizedFigureWithCaption#1#2#3{%
\setbox\ourfigbox
  \hbox{\hss\epsfxsize #1 \epsfbox{#2}\hss}
\hbox to \wd\ourfigbox{\vbox{\noindent\copy\ourfigbox\break
\vskip -6mm      \hbox to \wd\ourfigbox{\hss#3\hss}}}
}

\def\spa#1.#2{\left\langle#1\,#2\right\rangle}
\def\spb#1.#2{\left[#1\,#2\right]}
\def\lor#1.#2{\left(#1\,#2\right)}
\def\sand#1.#2.#3{%
\left\langle\smash{#1}{\vphantom1}^{-}\right|{#2}%
\left|\smash{#3}{\vphantom1}^{-}\right\rangle}
\def\sandpp#1.#2.#3{%
\left\langle\smash{#1}{\vphantom1}^{+}\right|{#2}%
\left|\smash{#3}{\vphantom1}^{+}\right\rangle}
\def\sandpm#1.#2.#3{%
\left\langle\smash{#1}{\vphantom1}^{+}\right|{#2}%
\left|\smash{#3}{\vphantom1}^{-}\right\rangle}
\def\sandmp#1.#2.#3{%
\left\langle\smash{#1}{\vphantom1}^{-}\right|{#2}%
\left|\smash{#3}{\vphantom1}^{+}\right\rangle}

\newcommand{\Bmp}[1]{\langle #1\rangle}

\begin{document}
\hfuzz 10 pt

%@twocolumnfalse\endcsname 

\ifpreprint
\noindent
\hfill Saclay/SPhT--T05/115
\hfill hep-th/0507292
\fi

\title{All-Multiplicity Amplitudes with Massive Scalars}

\author{Darren Forde} 
\author{David A. Kosower} 
\affiliation{Service de Physique Th\'eorique, 
CEA--Saclay\\ 
F--91191 Gif-sur-Yvette cedex, France
%\\{\tt David.Kosower@cea.fr}
}

\date{June 3, 2005}

\begin{abstract}

We compute two infinite series of tree-level amplitudes with a massive
scalar pair and an arbitrary number of gluons.  We provide results for
amplitudes where all gluons have identical helicity, and amplitudes
with one gluon of opposite helicity.
These amplitudes are useful for unitarity-based one-loop
calculations in nonsupersymmetric gauge theories generally, and 
QCD in particular.

\end{abstract}

\pacs{12.38.Bx, 12.38.-t \hspace{1cm}}

\maketitle

%\vskip2.0pc]

%% Section I %%%%%%%%%%%%%%%%%%%%%%%%%%%%%%%%%%%%%%%%%%%%%%%%%%%%%%%%%

\renewcommand{\thefootnote}{\arabic{footnote}}
\setcounter{footnote}{0}

%%%%%%%%%%%%%%%%%%%%%%%%%%%%%%%%%%%%%%%%%%%%%%%%%

\section{Introduction}
\label{IntroSection}

Explicit computations of Standard Model processes
will play an essential role in probing beyond it.
The short-distance experimental environment at hadron colliders requires
the computation of many processes involving QCD interactions.  
Tree-level computations do not suffice for these purposes.
The coupling in QCD is sufficiently large, and varies sufficiently
with scale, that even a basic quantitative understanding~\cite{GloverReview}
 requires
the computation of next-to-leading order corrections~\cite{NLOFormalisms}
 to cross sections.
One-loop amplitudes are of course an essential ingredient of such corrections.

Within the unitarity-based 
method~\cite{UnitarityMethod,MassiveUnitarity,UnitarityCalculations},
one can decompose one-loop QCD amplitudes into contributions 
corresponding to $\NeqFour$,
$\NeqOne$, and remaining contributions.  For amplitudes where all external
particles are gluons, this decomposition for the gluon-loop contributions
of color-ordered amplitudes 
takes the following form,
\begin{equation}
A_n = A_n^{\NeqFour} - 4 A_n^{\NeqOne} + A_n^{\rm scalar}.
\end{equation}
That is, the remaining contributions correspond to scalars
circulating in the loop.
(The contributions of quarks circulating in the loop can also be written
in terms of $A^{\NeqOne}$ and $A^{\rm scalar}$.)
The supersymmetric contributions can be 
computed by performing the cut algebra strictly in four dimensions, with
only the loop integrations computed in $D=4-2\e$ dimensions.  The `scalar'
contributions require that the cut algebra, and the corresponding tree 
amplitudes fed into the unitarity machinery, also be computed in $D$ 
dimensions~\cite{MassiveUnitarity,RationalUnitarity,DDimExamples,%
MoreDDimExamples}.  
At one
loop, computing a scalar loop in $D$ dimensions is equivalent to
computing a massive scalar loop, and then integrating over the mass with
an appropriate weighting.  

The computation of tree-level amplitudes with massive scalars is thus of
use in the unitarity method for computing 
massless loop amplitudes in nonsupersymmetric
gauge theories.  The simplest such amplitudes, with up to four gluons
of positive helicity, were computed by Bern, Dixon, Dunbar, and one of
the authors~\cite{RationalUnitarity}.
Recently, 
Badger, Glover, Khoze and~Svr\v{c}ek (BGKS) have given~\cite{MassiveRecursion}
 a set of on-shell recursion relations for amplitudes with massive scalars.
They used the recursion relations to compute all massive scalar amplitudes with
up to four external gluons.
These relations extend the tree-level on-shell recurrence relations of
Britto, Cachazo, Feng, and Witten~\cite{BCFRecursion,BCFW}.  
On-shell recursion relations have also been applied to 
tree-level amplitudes by Luo and Wen~\cite{LuoWen} and by 
Britto, Feng, Roiban, Spradlin and Volovich~\cite{BFRSV};
 to tree-level gravitational amplitudes
by Bedford, Brandhuber, Spence, and Travaglini~\cite{BBSTgrav},
and by Cachazo and Svr\v{c}ek~\cite{CSgrav}; and to massive vector and
quark amplitudes by Badger, Glover, and Khoze~\cite{MassiveVector}.
The concept has also been applied to certain loop
amplitudes by Bern, Dixon, and one of the authors~\cite{LoopOnShellRecursion};
and to the direct calculation of some integral coefficients by
Bern, Bjerrum-Bohr, Dunbar, and Ita~\cite{RecursiveCoefficients}.
These relations grew out of 
investigations~\cite{BCF7,NeqFourSevenPoint,BCFII,NeqFourNMHV,%
BCFcoplanar,RSVNewTree}
 motivated by Witten's 
topological twistor-string
description~\cite{WittenTopologicalString} (as a 
weak--weak dual) of the $\NeqFour$ supersymmetric
gauge theory.  The roots of this duality
lie in Nair's description~\cite{Nair} of 
the simplest gauge-theory scattering
amplitudes in terms of projective-space correlators.
For use as building blocks in loop amplitudes, we need analytic 
expressions for the massive-scalar amplitudes.
While the original tree-level recursive 
approach~\cite{Recursion,LightConeRecursion,DixonTASI}
pioneered by Berends and Giele is more efficient for purely numerical 
purposes, the on-shell approach seems better suited to providing analytic
formul\ae{}.  (We discuss the computational complexity of the Berends--Giele
recursion relations in the appendix.)
Bern, Dixon, and one of the authors have recently
presented~\cite{OneLoopRationalRecursion} a `unitarity-bootstrap' approach
which combines four-dimensional unitarity cuts with use of a recursion
relation for the rational terms.  The amplitudes we compute here should
be useful for proving the factorization properties in {\it complex\/}
momenta required for the recursion relations part of this approach.

In this paper, we will provide formul\ae{} for two classes of amplitudes
with two color-adjacent massive scalars and $n$ gluons, where the gluons all
share the same helicity, or where one gluon has opposite helicity.  
In the next section, we document the notation and conventions we use;
in \sect{AllPlusSection}, we derive a form for the all-plus
amplitude, $A_n(1_s,2^+,\ldots,(n-1)^+,n_s)$.  In \sect{LastMinusSection},
we derive an expression for the
 amplitude with a negative-helicity
gluon adjacent to a scalar leg, $A_n(1_s,2^+,\ldots,(n-1)^-,n_s)$,
and extend it to one negative-helicity gluon in an arbitrary position in
\sect{GeneralMinusSection}, followed by concluding remarks.

\section{Notation and On-Shell Recursion Relation}
\label{OnShellRecursionSection}

We will write our expressions for various amplitudes using the spinor-helicity
formalism.  The formalism makes use of spinor products.
We follow the conventions of the standard QCD literature,
so that
\begin{equation}
\spa{j}.{l} = \langle j^- | l^+ \rangle = \bar{u}_-(k_j) u_+(k_l)\,, 
\hskip 2 cm
\spb{j}.{l} = \langle j^+ | l^- \rangle = \bar{u}_+(k_j) u_-(k_l)\, ,
\label{spinorproddef}
\end{equation}
where $u_\pm(k)$ is a massless Weyl spinor with momentum $k$ and positive
or negative chirality~\cite{SpinorHelicity,TreeReview}. We take
all legs to be outgoing. The two spinor products are related,
with $\spb{i}.{j} = \sign(k_i^0 k_j^0)\spa{j}.{i}^*$
so that,
\begin{equation}
\spa{i}.{j} \spb{j}.{i} = 2 k_i \cdot k_j\,.
\end{equation}
(Note that the bracket product $\spb{i}.{j}$ differs by an overall sign
from that commonly used in twistor-space
studies~\cite{WittenTopologicalString} and also in ref.~\cite{MassiveRecursion}.) 

In the amplitudes we consider, we will also encounter sums of momenta,
\begin{equation}
K_{i\cdots j} = k_i+\cdots + k_j,
\end{equation}
but in this paper the indices should {\it not\/} be interpreted in a cyclic
manner; if $i>j$, $K_{i\cdots j} = 0$.

\def\spar#1..#2{\left\langle\!\left\langle#1\cdots#2\right\rangle\!\right\rangle}
Let us also define a notation for a sequential product of spinor
products,
\begin{equation}
\spar{j_1}..{j_2} = \spa{j_1}.{(j_1+1)} \spa{(j_1+1)}.{(j_1+2)} \cdots
                      \spa{(j_2-1)}.{j_2},
\end{equation}
and one for the mass-subtracted square of momentum sums,
\begin{equation}
L_{i\cdots j} = \sum_{a=i}^j \sum_{b=i,\neq a}^j k_a\cdot k_b,
\end{equation}
so that if $k_1^2 = m_s^2$, and $k_2,\ldots,k_j$ are massless,
for example, then $L_{1\cdots j} = K_{1\cdots j}^2 - m_s^2$.

The on-shell recursion relations make use of complex momenta, obtained
by shifting spinors corresponding to massless momenta.  (One can also shift
momenta of massive particles, but we will not need to do so in this paper.)
A $(j,l)$ shift is defined by, 
\begin{eqnarray}
&|j^-\rangle &\rightarrow |j^-\rangle - z|l^-\rangle \,,\nonumber\\
&|l^+\rangle &\rightarrow |l^+\rangle + z|j^+\rangle \,, 
\label{SpinorShift}
\end{eqnarray}
with the remaining spinors unchanged.  It
gives the following shift of the momenta,
\begin{eqnarray}
&k_j^\mu &\rightarrow \hat k_j \equiv k_j^\mu(z) = k_j^\mu - 
      {z\over2}{\sand{j}.{\gamma^\mu}.{l}},\nonumber\\
&k_l^\mu &\rightarrow k_l^\mu(z) = k_l^\mu + 
      {z\over2}{\sand{j}.{\gamma^\mu}.{l}}.
\label{MomentumShift}
\end{eqnarray}

To obtain an explicit expression using the on-shell recursion relations,
we must choose a definite shift.  The relations then express an amplitude
in terms of a sum over all inequivalent contiguous partitions of the external
momenta into two sets, each of which contain exactly one of the shift momenta.
These partitions can thus be thought of as corresponding to sums of
cyclicly-consecutive momenta.  There is also a sum over
the helicities of the intermediate leg.  The basic 
relation~\cite{BCFRecursion,BCFW,MassiveRecursion} is,
\def\Ph{{\hat P}}
\def\Pb{{\overline P}}
\def\Psl{\s{P}}
\begin{eqnarray}
A_n(k_1,\ldots,k_n) &=& \sum_{{\rm partitions}\, P} \sum_{h = \pm}
A_L(k_{P_1},\ldots,\hat k_j,\ldots,k_{P_{-1}},-\Ph^h)\nonumber\\
&&\hphantom{\sum\sum}\times
{i\over P^2-m_P^2}
A_R(k_{\Pb_1},\ldots,\hat k_l,\ldots,k_{\Pb_{-1}},\Ph^{-h}).
\end{eqnarray}
In this equation, $P_1$ stands for the first momentum in $P$, 
$P_{-1}$ for the last momentum in $P$, and $\Pb$ for the complement of
$P$ (all the remaining momenta).  The scalar legs
(always the first and last in this paper) will be indicated by an `$s$'
subscript.  The mass of the particle in the factorized
channel is $m_P$, and $\Ph$ is given by momentum conservation,
\begin{equation}
\Ph = P + \delta k_j(z) = k_{P_1}+\cdots + k_{P_{-1}} 
                 - {z\over2}{\sand{j}.{\gamma^\mu}.{l}}.
\end{equation}
In each channel, a different value of $z$ is used here, and in 
\eqns{SpinorShift}{MomentumShift}.  It is given by the on-shell condition
$\Ph^2 - m_P^2 = 0$,
\begin{equation}
z = {P^2 - m_P^2\over \sand{j}.{\Psl}.{l}}.
\end{equation}

As discussed in refs.~\cite{BCFW,MassiveRecursion}, 
we must ensure that $A(z)$ vanishes at large
$z$ for the recursion relations to be valid.  The choices we will use
have the required property, as was shown in these references.

\section{The $A(1_s,2^+,\ldots,(n-1)^+,n_s)$ Amplitude}
\label{AllPlusSection}

\def\LP{\left.}
\def\LB{\left[}
\def\RB{\right]}
\def\RV{\right|}
\def\nmo{(n\!-\!1)}

In this section, we provide a result for the amplitude with an
arbitrary number of positive-helicity gluons.  The amplitude with all
negative-helicity gluons can be obtained by spinor conjugation.  Let us
begin at the end, by writing down the answer,
\begin{eqnarray}
&&A_{n}(1_s,2^+,\ldots,\nmo^+,n_s) = 
-{i\over L_{12} \spar2..{\nmo} L_{(n-1)n}}\nonumber\\
&&\times\mskip -10mu
 \sum_{j=1}^{\lfloor n/2\rfloor-1} \mskip -7mu (-m_s^{2})^{j}\mskip -12mu
    \sum_{\{w_i\}_{i=1}^{j-1}\atop w_i=w_{i-1}+2}^{n-3}
      \mskip -5mu
       \biggl(\prod_{r=1}^{j-1} 
              {\sand{w_r}.{\Ksl_{(w_r+1)\cdots (w_{r+1}-1)}}.{w_{r+1}}
               \over L_{1\cdots (w_r-1)} L_{1\cdots w_r}}\biggr)
          \sandpm{w_1}.{\Ksl_{2\cdots (w_1-1)}\ksl_1}.2
    \LP\vphantom{\sum_{i=1}^{n-3}}
    \RV_{w_0=1,\atop w_j=n-1}
\label{AllPlusResult}
\end{eqnarray}
In this equation, $m_s^2 \equiv k_1^2 = k_n^2$ is the mass squared of
the scalar, and $\lfloor x\rfloor$ denotes the largest integer smaller than or
equal to $x$.

The multiple sums in \eqn{AllPlusResult} can be written less succinctly 
and perhaps less forbiddingly as,
\begin{equation}
\sum_{\{w_i\}_{i=1}^{j-1}\atop w_i=w_{i-1}+2}^{n-3}
\LP\vphantom{ \sum_{w_1=2}^{n-3} }\RV_{w_0=1,\atop w_j=n-1}
= \LP\sum_{w_1=3}^{n-3} \sum_{w_2=w_1+2}^{n-3} \cdots \sum_{w_{j-1}=w_{j-2}+2}^{n-3}
\RV_{w_j=n-1}.
\end{equation}

For example, in the four-point case, the product over $r$ is absent,
and we obtain,
\begin{equation}
{i m_s^2 \spb3.2 \sand2.{\ksl_1}.2\over L_{12} \spa2.3 L_{34}} =
%%%%% begin : ampl4spps
{i m_s^2 \spb3.2 \over L_{12} \spa2.3};
%%%%% end : ampl4spps
\end{equation}
in the five-point case, 
\begin{equation}
%%%%% begin : ampl5sppps
{i m_s^2\sandpm4.{\Ksl_{23}\ksl_1}.2\over L_{12} \spa2.3 \spa3.4 L_{45}}
%%%%% end : ampl5sppps
\end{equation}
and in the six-point case, where there is only one non-trivial $w$ variable,
\begin{equation}
%%%%% begin : ampl6spppps
{i m_s^2\over L_{12} \spa2.3\spa3.4\spa4.5 L_{56}}
\biggl( \sandpm5.{\Ksl_{2\cdots4}\ksl_1}.2
-m_s^2 {\sandpm{5}.{\ksl_4\ksl_{3}}.2
                    \over L_{4\cdots 6}}
%- m_s^2 \sum_{w=2}^3 {\sand{w}.{\Ksl_{(w+1)\cdots 4}}.5
%                    \sandpm{w}.{\Ksl_{2\cdots (w-1)}\ksl_1}.2
%                    \over L_{1\cdots (w-1)} L_{(w+1)\cdots 6}}
\biggr)
%%%%% end : ampl6spppps
\end{equation}
all in agreement with refs.~\cite{RationalUnitarity,MassiveRecursion}
(up to an overall phase for the second reference).

\def\Kh{{\hat K}}
\def\Khsl{\s{\Kh}}
\def\khsl{\s{{\hat k}}}
Having written down the above ansatz, we now prove it using the on-shell
recursion relations as given by BGKS.  For this purpose, let us choose
a $(3,2)$ shift using \eqns{SpinorShift}{MomentumShift}.
(That is, choose the `reference' momenta to be $k_{3}$ and $k_2$.)
\iffalse  The
corresponding massless spinors are shifted 
as follows,
\begin{eqnarray}
&\tlambda_3 &\rightarrow \tlambda_3 - z\tlambda_2 \,,\nonumber\\
&\lambda_2 &\rightarrow \lambda_2 + z\lambda_3 \,, 
\label{SpinorShift}
\end{eqnarray}
and the momenta,
\begin{eqnarray}
&k_2^\mu &\rightarrow k_2^\mu(z) = k_2^\mu + 
      {z\over2}{\sand{3}.{\gamma^\mu}.{2}},\nonumber\\
&k_3^\mu &\rightarrow k_3^\mu(z) = k_3^\mu - 
      {z\over2}{\sand{3}.{\gamma^\mu}.{2}},
\label{MomentumShift}
\end{eqnarray}
\fi

In \eqn{AllPlusResult}, the limits on the sums ensure that $i\leq j$
in sums of momenta $K_{i\cdots j}$.  We can remove those limits, so long
as we take $K_{i\cdots j}\equiv 0$ if $i>j$ (rather than interpreting the
indices in a cyclic sense), and we shall do so.  We can then rewrite
\eqn{AllPlusResult} in a form which will be more useful for the proof,
\begin{eqnarray}
&&A_{n}(1_s,2^+,\ldots,\nmo^+,n_s) = 
-{i\over L_{12} \spar2..{\nmo} L_{(n-1)n}}\nonumber\\
&&\times\mskip -10mu
 \sum_{j=1}^{\lfloor n/2\rfloor-1} \mskip -7mu (-m_s^{2})^{j}\mskip -12mu
    \sum_{\{w_i\}_{i=1}^{j-1}=3}^{n-3}
      \mskip -5mu
       \biggl(\prod_{r=1}^{j-1} 
              {\sand{w_r}.{\Ksl_{(w_r+1)\cdots (w_{r+1}-1)}}.{w_{r+1}}
               \over L_{1\cdots (w_r-1)} L_{1\cdots w_r}}\biggr)
          \sandpm{w_1}.{\Ksl_{2\cdots (w_1-1)}\ksl_1}.2
    \LP\vphantom{\sum_{i=1}^{n-3}}
    \RV_{w_j=n-1}
\label{AllPlusResult2}
\end{eqnarray}

We will proceed inductively, assuming that the ansatz~(\ref{AllPlusResult})
holds for the $(n-1)$-point amplitude ($n>4$), and showing that it holds for the
$n$-point one.  The on-shell recursion 
relations~\cite{BCFRecursion,BCFW,MassiveRecursion} tell us that the $n$-point
amplitude can be written as a sum over all factorizations, with the two
shifted momenta attached to different amplitudes in each factorization.

No factorization can isolate $k_2$ in a purely-gluonic
amplitude.  In addition,
any factorization that isolates $k_3$ in a purely-gluonic
amplitude leads to a vanishing contribution because the amplitude vanishes.
For factorizations where $k_3$ is isolated in a four- or higher-point amplitude,
this is immediate, because all such amplitudes vanish,
\begin{equation}
A_{j+1}(\Kh_{3\cdots j}^\pm,3^+,\ldots,j^+) = 0.
\end{equation}
In the three-point case, the factorization is in the $\spb3.4$ channel,
but here the relevant amplitude --- 
$A_{3}(\Kh_{34}^\pm,3^+,4^+)$
--- also vanishes.

This leaves us only with factorization in which both amplitudes have
massive scalar legs, and therefore in which the factorized leg is 
a massive scalar.  There is only one such contribution, so what we are
seeking to prove is that,
\begin{equation}
A_n(1_s,2^+,\ldots,(n-1)^+,n_s) = 
A_3(1_s,2^+,-\Kh_{12s}) {i\over K_{12}^2-m_s^2}
A_{n-1}(\Kh_{12s},3^+,\ldots,(n-1)^+,n_s).
\label{AllPlusStartingEquation}
\end{equation}

The three-point amplitude was given by BGKS~\cite{MassiveRecursion},
\begin{equation}
A_3(1_s,2^+,3_s) = i{\sand{q}.{\ksl_1}.2\over\spa{q}.2}.
\end{equation}
Its form depends on an arbitrary reference momentum $q$, but its value
is nonetheless independent of it on shell.

Using this expression with $q=\hat k_3$, and rewriting denominators
using momentum conservation, we have for our starting point,
\def\spacer{\hphantom{\spa1.2}}
\begin{eqnarray}
&&-{i 
    \over L_{12} L_{123} \spar2..{\nmo} L_{(n-1)n}}\nonumber\\
&&\spacer\times\mskip -15mu
 \sum_{j=1}^{\lfloor (n-1)/2\rfloor-1} 
        \mskip -15mu (-m_s^{2})^{j}\mskip -10mu
    \sum_{\{w_i\}_{i=1}^{j-1}=4}^{n-3}
      \mskip -5mu
       \biggl(\prod_{r=1}^{j-1} 
              {\sand{w_r}.{\Ksl_{(w_r+1)\cdots (w_{r+1}-1)}}.{w_{r+1}}
               \over L_{w_r\cdots n} L_{(w_r+1)\cdots n}}\biggr)
    \nonumber\\
&&\spacer \hphantom{\sand{w_r}.{\Ksl_{(w_r+1)\cdots (w_{r+1}-1)}}.{w_{r+1}} }%
  \times
    \LP\vphantom{\sum_{i=1}^{n-3}}
  \sandpm{w_1}.{\Ksl_{\hat 3\cdots (w_1-1)}\Khsl_{12}\khsl_3\ksl_1}.2
   \RV_{w_j=n-1},
\end{eqnarray}
where we have also used the fact that 
$|\hat 2^-\rangle = |2^-\rangle$ and
$\langle \hat 3^-| = \langle 3^-|$.
Now,
\begin{eqnarray}
\Khsl_{12}\khsl_3\ksl_1|2^-\rangle
&=& 2 \Kh_{12}\cdot k_3\, \ksl_1|2^-\rangle
-\khsl_3\Khsl_{12}\ksl_1|2^-\rangle\nonumber\\
&=& L_{123}\, \ksl_1|2^-\rangle
-m_s^2\khsl_3|2^-\rangle,
\end{eqnarray}
thanks to the on-shell conditions in the three-point amplitude.

Writing out the two terms, we have
\begin{eqnarray}
&&-{i
    \over L_{12} \spar2..{\nmo} L_{(n-1)n}}
   \left\{\vphantom{ \sum_{\{w_i\}_{i=1}^{j-1}=4}^{n-3} }\right.\nonumber\\
&&\mskip -10mu
 \sum_{j=1}^{\lfloor (n-1)/2\rfloor-1}
   \mskip -20mu (-m_s^{2})^{j} \mskip -12mu
    \sum_{\{w_i\}_{i=1}^{j-1}=4}^{n-3}
      \mskip -5mu
       \biggl(\prod_{r=1}^{j-1} 
              {\sand{w_r}.{\Ksl_{(w_r+1)\cdots (w_{r+1}-1)}}.{w_{r+1}}
               \over L_{w_r\cdots n} L_{(w_r+1)\cdots n}}\biggr)
  \sandpm{w_1}.{\Ksl_{\hat 3\cdots (w_1-1)}\ksl_1}.2
   \LP\vphantom{\sum_{i=1}^{n-3}}
   \RV_{w_j=n-1} \label{TwoTerms}\\
&&\mskip -10mu
 +\mskip -10mu \sum_{j=2}^{\lfloor (n-1)/2\rfloor} 
   \mskip -12mu (-m_s^{2})^{j} \mskip -12mu
    \sum_{\{w_i\}_{i=2}^{j-1}=4}^{n-3}
      \mskip -5mu
       \biggl(\prod_{r=2}^{j-1} 
              {\sand{w_r}.{\Ksl_{(w_r+1)\cdots (w_{r+1}-1)}}.{w_{r+1}}
               \over L_{w_r\cdots n} L_{(w_r+1)\cdots n}}\biggr)
  {\sandpm{w_2}.{\Ksl_{3\cdots (w_2-1)}\ksl_3}.2
   \over L_{4\cdots n}}
   \LP\vphantom{\sum_{i=1}^{n-3}}
   \RV_{w_j=n-1} 
     \left.\vphantom{ \sum_{\{w_i\}_{i=1}^{j-1}=4}^{n-3} }\right\}.
\nonumber
\end{eqnarray}
We can rewrite the last factor in brackets,
\begin{eqnarray}
{\sandpm{w_2}.{\Ksl_{3\cdots (w_2-1)}\ksl_3}.2
   \over L_{4\cdots n}} &=&
\LP{\sand{w_1}.{\Ksl_{4\cdots (w_2-1)}}.{w_2}
   \over L_{w_1\cdots n} L_{(w_1+1)\cdots n}}
   {\sandpm{w_1}.{\Ksl_{2\cdots (w_1-1)}\ksl_1}.2}\RV_{w_1=3}.
\end{eqnarray}
We also note that because of the on-shell conditions on the three-point
amplitude, ${2 \hat k_2\cdot k_1 = 0 = \sand{\hat 2}.{\ksl_1}.{2}}$, so
that we can replace $\sandpm{w_1}.{\Ksl_{\hat 3\cdots (w_1-1)}\ksl_1}.2$
in the last factor by $\sandpm{w_1}.{\Ksl_{2\cdots (w_1-1)}\ksl_1}.2$.

Furthermore, we can extend all but the $w_1$ sums in \eqn{TwoTerms} down
to $w=3$, since the summands will necessarily vanish.  The second term
will then supply the $w_1=3$ terms, with some left-over pieces,
\begin{eqnarray}
&&-{i 
    \over L_{12} \spar2..{\nmo} L_{(n-1)n}}
   \left\{\vphantom{ \sum_{\{w_i\}_{i=1}^{j-1}=4}^{n-3} }\right.\nonumber\\
&&
 \sum_{j=1}^{\lfloor (n-1)/2\rfloor-1}
   \mskip -12mu (-m_s^{2})^{j} \mskip -12mu
    \sum_{\{w_i\}_{i=1}^{j-1}=3}^{n-3}
      \mskip -5mu
       \biggl(\prod_{r=1}^{j-1} 
              {\sand{w_r}.{\Ksl_{(w_r+1)\cdots (w_{r+1}-1)}}.{w_{r+1}}
               \over L_{w_r\cdots n} L_{(w_r+1)\cdots n}}\biggr)
  \sandpm{w_1}.{\Ksl_{2\cdots (w_1-1)}\ksl_1}.2
   \LP\vphantom{\sum_{i=1}^{n-3}}
   \RV_{w_j=n-1}\nonumber\\
&&
 +(-m_s^{2})^{j}
    \sum_{\{w_i\}_{i=2}^{j-1}=3}^{n-3}
      \mskip -5mu
       \biggl(\prod_{r=1}^{j-1} 
              {\sand{w_r}.{\Ksl_{(w_r+1)\cdots (w_{r+1}-1)}}.{w_{r+1}}
               \over L_{w_r\cdots n} L_{(w_r+1)\cdots n}}\biggr)
   {\sandpm{w_1}.{\Ksl_{2\cdots (w_1-1)}\ksl_1}.2}
   \LP\vphantom{\sum_{i=1}^{n-3}}
   \RV_{j=\lfloor (n-1)/2\rfloor\atop w_1=3, w_j=n-1} \nonumber\\
&&
 -(-m_s^{2})^{j}
    \sum_{\{w_i\}_{i=2}^{j-1}=3}^{n-3}
      \mskip -5mu
   {\sandpm{w_1}.{\Ksl_{2\cdots (w_1-1)}\ksl_1}.2}
   \LP\vphantom{\sum_{i=1}^{n-3}}
   \RV_{j=1\atop w_1=3, w_j=n-1}
     \left.\vphantom{ \sum_{\{w_i\}_{i=1}^{j-1}=4}^{n-3} }\right\}.
\label{TwoTerms2}
\end{eqnarray}
The last term is absent for $n>4$ (because the constraints are incompatible). 
 For the penultimate term, we consider the
even and odd cases separately.

If $n$ is even,
\begin{equation}
 \sum_{j=1}^{\lfloor (n-1)/2\rfloor-1} =
 \sum_{j=1}^{\lfloor n/2\rfloor-2},
\end{equation}
and the penultimate term completes the sum in the first term to
up to $j=\lfloor n/2\rfloor-1$.  Note that in the penultimate term, $w_1$ can in
any event take only the value $3$, because while we have $w_r\geq 2r+1$,
the presence of $n/2-2$ different summation indices $w_r$ with 
$w_r \geq w_{r-1}+2$ requires
that $w_r$ take {\it exactly} the value $2r+1$.

If $n$ is odd, on the other hand, then the penultimate term in \eqn{TwoTerms2}
vanishes, because we have $(n-3)/2$ different summation indices $w_r$, and
the last one would have to obey the incompatible constraints 
$w_{(n-3)/2} \leq n-3$ and $w_{(n-3)/2} \geq n-2$.

We thus obtain \eqn{AllPlusResult2}, as desired.  We have verified this
equation
numerically, using a set of light-cone recursion relations of
the conventional kind~\cite{LightConeRecursion}, through $n=12$.

\section{The $A(1_s,2^+,\ldots,(n-2)^+,(n-1)^-,n_s)$ Amplitude}
\label{LastMinusSection}

We turn next to an amplitude with one negative helicity, adjacent to
a scalar leg.  There are two such color-ordered amplitudes, related by
reflection.

Here, we shall use an $(n-1,n-2)$ shift,
\begin{eqnarray}
&|(n-1)^-\rangle &\rightarrow |(n-1)^-\rangle - z|(n-2)^-\rangle \,,\nonumber\\
&|(n-2)^+\rangle &\rightarrow |(n-2)^+\rangle + z|(n-1)^+\rangle \,, 
\label{LastMinusSpinorShift}
\end{eqnarray}

The recursion for the target amplitude now has two terms,
\def\spacer{\hskip 5mm}
\begin{eqnarray}
&&A_n(1_s,2^+,\ldots,{\widehat {(n-2)}}{}^+,
      {\widehat {(n-1)}}{}^-,n_s) =\nonumber\\
&&\spacer
A_{n-1}(1_s,2^+,\ldots,{\widehat {(n-2)}}{}^+,\Kh_{(n-1)n\,s}^-) 
{i\over K_{(n-1)n}^2-m_s^2} 
A_3(-\Kh_{(n-1)n\,s}^+,{\widehat {(n-1)}}{}^-,n_s)
\nonumber\\
&&\spacer
+A_{n-1}(1_s,2^+,\ldots,\Kh_{(n-3)(n-2)}^+,{\widehat {(n-1)}}{}^-,n_s) 
\label{LastMinusStartingEquation}\\
&&\spacer\hphantom{A_{n-1}()}\times
{i\over K_{(n-3)(n-2)}^2} 
A_3(-\Kh_{(n-3)(n-2)}^-,(n-3)^+,{\widehat {(n-2)}}{}^+).\nonumber
\end{eqnarray}

\def\spacer{\hskip 2mm}
\def\lshift{\hskip -20pt}
\def\spash#1.#2{\vphantom{\hat K#1#2}\spa{\smash{#1}}.{\smash{#2}}}
\def\spbsh#1.#2{\vphantom{\hat K#1#2}\spb{\smash{#1}}.{\smash{#2}}}
We can evaluate the first term directly,
\begin{eqnarray}
&& \lshift T_1(1_s,2^+,\ldots,{\widehat {(n-2)}}{}^+,{\widehat {(n-1)}}{}^-,n_s) \equiv
  \nonumber\\
&&\lshift\spacer\hphantom{=!}
A_{n-1}(1_s,2^+,\ldots,{\widehat {(n-2)}}{}^+,\Kh_{(n-1)n\,s}^-) 
{i\over K_{(n-1)n}^2-m_s^2} 
A_3(-\Kh_{(n-1)n\,s}^+,{\widehat {(n-1)}}{}^-,n_s)\nonumber\\
&&\lshift\spacer= -{i\over L_{12} \spar1..{(n-3)} \spash{(n-3)}.{\widehat {(n-2)}}
             L_{{\widehat {(n-2)}}\Kh_{(n-1)n}} L_{{\widehat {(n-1)}}n}}
   {\sand{{\widehat {(n-1)}}}.{\Khsl_{(n-1)n}}.q\over\spbsh{q}.{{\widehat {(n-1)}}}}
\\
&&\lshift\spacer\hphantom{=!}\hskip 3mm\times
      \mskip -15mu
 \sum_{j=1}^{\lfloor (n-1)/2\rfloor-1} 
      \mskip -20mu (-m_s^{2})^{j} \mskip -12mu
    \sum_{\{w_i\}_{i=1}^{j-1}=3}^{n-4}
      \mskip -5mu
       \biggl(\prod_{r=1}^{j-1} 
              {\sand{w_r}.{\Ksl_{(w_r+1)\cdots (w_{r+1}-1)}}.{w_{r+1}}
               \over L_{1\cdots (w_r-1)} L_{1\cdots w_r}}\biggr)
          \sandpm{w_1}.{\Ksl_{2\cdots (w_1-1)}\ksl_1}.2
    \LP\vphantom{\sum_{i=1}^{n-3}}
    \RV_{w_j=n-1}\nonumber\\
&&\lshift\spacer = -{i\,\sand{(n-1)}.{\ksl_n}.{(n-2)}^2\over
      L_{12} \spar2..{(n-3)} 
      \sand{(n-3)}.{\Ksl_{(n-2)(n-1)}\ksl_n\Ksl_{(n-2)(n-1)}}.{(n-2)}
      L_{(n-2)\cdots n} L_{(n-1)n}}\nonumber\\
&&\lshift\spacer\hphantom{=!}\hskip 3mm\times
      \mskip -15mu
 \sum_{j=1}^{\lfloor (n-1)/2\rfloor-1}
      \mskip -20mu (-m_s^{2})^{j} \mskip -12mu
    \sum_{\{w_i\}_{i=1}^{j-1}=3}^{n-4}
      \mskip -5mu
       \biggl(\prod_{r=1}^{j-1} 
              {\sand{w_r}.{\Ksl_{(w_r+1)\cdots (w_{r+1}-1)}}.{w_{r+1}}
               \over L_{1\cdots (w_r-1)} L_{1\cdots w_r}}\biggr)
          \sandpm{w_1}.{\Ksl_{2\cdots (w_1-1)}\ksl_1}.2
    \LP\vphantom{\sum_{i=1}^{n-3}}
    \RV_{w_j=n-1}\nonumber
\end{eqnarray}

For the second term, 
\begin{eqnarray}
&&T_2(1_s,2^+,\ldots,{\widehat {(n-2)}}{}^+,{\widehat {(n-1)}}{}^-,n_s) \equiv
  \nonumber\\
&&\spacer A_{n-1}(1_s,2^+,\ldots,\Kh_{(n-3)(n-2)}^+,{\widehat {(n-1)}}{}^-,n_s) 
\\
&&\spacer\hphantom{A_{n-1}()} \times {i\over K_{(n-3)(n-2)}^2} 
A_3(-\Kh_{(n-3)(n-2)}^-,(n-3)^+,{\widehat {(n-2)}}{}^+),
\nonumber\end{eqnarray}
we can iterate the recursion relations; suppressing all arguments but
the number of legs, we obtain the following structure,
\begin{eqnarray}
A_n &=& T_1(n) + {i A_3 \over P^2} A_{n-1}\nonumber\\
&=& T_1(n) + {i A_3\over P^2} \Bigl[ T_1(n-1) + {i A_3\over P^2} A_{n-2}\Bigr]
\label{Iteration}\\
&=& T_1(n) + {i A_3\over P^2} \biggl[ T_1(n-1) + {i A_3\over P^2} 
                          \Bigl[T_1(n-2) + {i A_3\over P^2} A_{n-3}\Bigr]\biggr]
\nonumber\\
&\vdots&\nonumber
\end{eqnarray}

\def\dKh{\dblhat{K}}
In this iteration, the $\Kh$ momentum at any stage will be one of the shifted
momenta at the following stage; to make this explicit, it will be helpful to
define,
\begin{eqnarray}
\Kh_{[1]}^\mu &=& k_{m-1}^\mu,\nonumber\\
\Kh_{[j]}^\mu &=& \Kh_{m-j,\Kh_{[j-1]}}=k_{m-j}^\mu + \Kh_{[j-1]}^\mu 
 - {\bigl(k_{m-j}+ \Kh_{[j-1]}\bigr)^2 \sand{m}.{\gamma^\mu}.{\Kh_{[j-1]}}
               \over 2 \sand{m}.{\ksl_{m-j}}.{\Kh_{[j-1]}}},
\label{eq:shifted_K_mom_def}\\
\dKh_{[j]}^{\mu}&=&\Kh_{[j]}^{\mu}
-\frac{(k_{m-j-1}+\hat{K}_{[j]})^2
  \sand{m}.{\gamma^{\mu}}.{\Kh_{[j]}}}{2\sand{m}.{\ksl_{m-j-1}}.{\Kh_{[j]}}},
\nonumber
\end{eqnarray}
so that
$\dKh_{[j-1]}+k_{m-j}-\Kh_{[j-1]}=0$.  
For the specific configuration considered
 in this section $m=n-1$, and hence we have
\begin{eqnarray}
\Kh_{[1]}^\mu &=& k_{n-2}^\mu,\nonumber\\
\Kh_{[j]}^\mu &=& k_{n-j-1}^\mu + \Kh_{[j-1]}^\mu 
 - {\bigl(k_{n-j-1}+ \Kh_{[j-1]}\bigr)^2 \sand{(n-1)}.{\gamma^\mu}.{\Kh_{[j-1]}}
               \over 2 \sand{(n-1)}.{\ksl_{n-j-1}}.{\Kh_{[j-1]}}},
\\ \dKh_{[j]}^{\mu}&=&\Kh_{[j]}^{\mu}
-\frac{(k_{n-j-2}+\hat{K}_{[j]})^2
  \sand{(n-1)}.{\gamma^{\mu}}.{\Kh_{[j]}}}{2\sand{(n-1)}.{\ksl_{n-j-2}}.{\Kh_{[j]}}}.
\nonumber
\end{eqnarray}

Putting the arguments back into \eqn{Iteration}, 
%and adding two additional arguments to $A_j$ (separated by a semicolon)
%to specify the shift spinors, 
we obtain an explicit expression
for $A_n$,
\def\spacer{\hskip 5mm}
\begin{eqnarray}
&&A_n(1_s,2^+,\ldots, {(n-2)}^+,{(n-1)}^-,n_s) =
  \nonumber\\
&&\spacer \sum_{j=1}^{n-3}
\prod_{r=2}^{j} {i A_3 (-\Kh_{[r]}^-,(n-1-r)^+,\dKh_{[r-1]}^+)
                 \over K_{(n-2-r)\cdots(n-2)}^2}
\nonumber\\
&&\spacer \hskip 4mm\times T_1(1_s,2^+,\ldots,(n-2-j)^+,\Kh_{[j]}^+,
                               {\widehat{(n-1)}}{}^-,n_s).
\end{eqnarray}

To proceed,
we must simplify the product of three-point amplitudes.  We can first
rewrite
\begin{eqnarray}
&&{i A_3 (-\Kh_{[2]}^-,(n-3)^+,\dKh_{[1]}^+)
                 \over K_{(n-3)(n-2)}^2} =\nonumber\\
&&\hskip 10mm
\frac{\spbsh{(n-3)}.{\dKh_{[1]}}^4}
{\spbsh{(-\Kh_{[2]})}.{(n-3)}
\spbsh{(n-3)}.{\dKh_{[1]}}
\spbsh{\dKh_{[1]}}.{(-\Kh_{[2]})}K_{(n-3)(n-2)}^2}
\nonumber\\
&&\hskip 10mm
= -{\spash{(n-1)}.{(-\Kh_{[2]})}^2
 \over \spa{(n-1)}.{(n-3)} \spa{(n-3)}.{(n-2)} \spa{(n-2)}.{(n-1)}},
\end{eqnarray}
and then by induction, we can show that
\begin{eqnarray}
&&\prod_{r=2}^{j} {iA_3 (-\Kh_{[r]}^-,(m-r)^+,\dKh_{[r-1]}^+)
                 \over K_{(m-r)\cdots(m-1)}^2}
=\nonumber\\
&&\hphantom{\prod_{r=2}^{j}}\hskip 10mm
 -{\spash{m}.{(-\Kh_{[j]})}^2
   \over \spa{m}.{(m-j+1)} \spar{(m-j+1)}..{m}}.
\label{eq:A_3_product_result}
\end{eqnarray}
where for the case considered in this section $m=n-1$.

Using this result, we obtain our final formula for the
amplitude,
\begin{eqnarray}
&&A(1_s,2^+,\ldots,(n-2)^+,(n-1)^-,n_s) =\nonumber\\
&&\spacer {i\sandmp{(n-1)}.{\ksl_1\ksl_n}.{(n-1)}^2
           \over \spar2..{(n-1)} K_{2\ldots (n-1)}^2 
                 \sandmp{(n-1)}.{\Ksl_{2\ldots(n-1)}\ksl_1}.2}\nonumber\\
&&\spacer+\sum_{l=1}^{n-4}
 {i\spa{(n-l-2)}.{(n-l-1)}
  \over\spar2..{(n-1)} L_{12} K_{(n-l-1)\cdots (n-1)}^2 
       L_{(n-l-1)\cdots n}^{\vphantom2}}
\label{LastMinusResult}\\
&&\spacer 
 \times {\sandmp{(n-1)}.{\Ksl_{(n-l-1)\cdots(n-1)}\ksl_n}.{(n-1)}^2
  \over\sandmp{(n-l-2)}.{\Ksl_{(n-l-1)\cdots (n-1)}\ksl_n}.{(n-1)}
       \sandmp{(n-l-1)}.{\Ksl_{(n-l-1)\cdots (n-1)}\ksl_n}.{(n-1)} }
\nonumber\\
&&\spacer\hskip 5mm\times
  \mskip -10mu
 \sum_{j=1}^{\lfloor (n-l)/2\rfloor-1} \mskip -7mu (-m_s^{2})^{j}\mskip -12mu
    \sum_{\{w_i\}_{i=1}^{j-1}\atop w_i=w_{i-1}+2}^{n-l-3}
      \mskip -5mu
       \biggl(\prod_{r=1}^{j-1} 
              {\sand{w_r}.{\Ksl_{(w_r+1)\cdots (w_{r+1}-1)}}.{w_{r+1}}
               \over L_{1\cdots (w_r-1)} L_{1\cdots w_r}}\biggr)
\nonumber\\
&&\spacer\hskip 5mm%
 \hphantom{ \sand{w_r}.{\Ksl_{(w_r+1)\cdots (w_{r+1}-1)}}.{w_{r+1}} }\times
          \sandpm{w_1}.{\Ksl_{2\cdots (w_1-1)}\ksl_1}.2
    \LP\vphantom{\sum_{i=1}^{n-3}}
    \RV_{\langle w_j^+| = \langle (n-1)^-|\Ksls_{(n-l-1)\cdots (n-1)},
         \atop w_0=1, w_j=n-l-1}
\nonumber
\end{eqnarray}
where we have separated out the $m$-independent terms.  The reader can
readily verify that these reproduce the required maximally
helicity-violating (MHV) amplitude in the
massless limit.

We can rewrite \eqn{LastMinusResult} in a slightly more compact form
as
\begin{eqnarray}
&&A(1_s,2^+,\ldots,(n-2)^+,(n-1)^-,n_s) =\nonumber\\
%%%%% begin : ampln1malt1
&&\spacer {i\sandmp{(n-1)}.{\ksl_1\ksl_n}.{(n-1)}^2
           \over \spar2..{(n-1)} K_{2\cdots (n-1)}^2 
                 \sandmp{(n-1)}.{\Ksl_{2\cdots(n-1)}\ksl_1}.2}\nonumber\\
%%%%% end : ampln1malt1
&&\spacer+\sum_{l=1}^{n-4}
%%%%% begin : ampln1malt2summand
 {i\spa{(n-l-2)}.{(n-l-1)}
  \over\spar2..{(n-1)} L_{12} K_{(n-l-1)\cdots (n-1)}^2 
       L_{(n-l-1)\cdots n}^{\vphantom2}}
\label{LastMinusResult2}\\
&&\spacer \times 
{\sandmp{(n-1)}.{\Ksl_{(n-l-1)\cdots(n-1)}\ksl_n}.{(n-1)}^2 G(2,n-l-1;n-1)
  \over\sandmp{(n-l-2)}.{\Ksl_{(n-l-1)\cdots (n-1)}\ksl_{n}}.{(n-1)}
       \sandmp{(n-l-1)}.{\Ksl_{(n-l-1)\cdots (n-1)}\ksl_{n}}.{(n-1)} }
%%%%% end : ampln1malt2summand
\nonumber
\end{eqnarray}
by introducing
\begin{eqnarray}
G(a_1,b_1;m)&=&\sum_{j=1}^{\lfloor (b_1-a_1+1)/2\rfloor} \mskip -7mu (-m_s^{2})^{j}\mskip -12mu
    \sum_{\{w_i\}_{i=1}^{j-1}\atop w_i=w_{i-1}+2}^{b_1-2}
      \mskip -5mu
       \biggl(\prod_{r=1}^{j-1} 
              {\sand{w_r}.{\Ksl_{(w_r+1)\cdots (w_{r+1}-1)}}.{w_{r+1}}
               \over L_{1\cdots (w_r-1)} L_{1\cdots w_r}}\biggr)
\nonumber\\
&&\spacer\hskip 5mm%
 \times \sandpm{w_1}.{\Ksl_{a_1\cdots (w_1-1)}\ksl_1}.{a_1}
    \LP\vphantom{\sum_{i=1}^{n-3}}
    \RV_{\langle w_j^+| = \langle m^-|\Ksls_{b_1\cdots m},
         \atop w_0=a_1-1, w_j=b_1}
\label{eq:def_simple_G}
\end{eqnarray}
%\hphantom{ \sand{w_r}.{\Ksl_{(w_r+1)\cdots (w_{r+1}-1)}}.{w_{r+1}} }

In the four-point case, only the first term in \eqn{LastMinusResult} is
present, and we obtain,
\begin{eqnarray}
A_4(1_s,2^+,3^-,4_s) &=& 
%%%%% begin : ampl4spms
        i {\sandmp3.{\ksl_1\ksl_4}.3^2
                         \over \spa2.3 K_{23}^2 \sandmp3.{\ksl_2\ksl_1}.2}
%%%%% end : ampl4spms
\nonumber\\
&=& -i{\sand3.{\ksl_1}.2^2\over K_{23}^2 L_{12}^{\vphantom2} };
\end{eqnarray}
in the five-point case,
\begin{eqnarray}
%%%%% begin : ampl5sppms
-{i\sandmp4.{\ksl_1\Ksl_{23}}.4^2\over
        K_{234}^2 \spa2.3\spa3.4 \sandmp2.{\ksl_1\Ksl_{23}}.4}
-{i m_s^2 \sand4.{\ksl_5}.3^2 \spb2.3
  \over L_{12} L_{45} \spb3.4 \sandmp4.{\ksl_5\Ksl_{34}}.2};
%%%%% end : ampl5sppms
\end{eqnarray}
and in the six-point case,
\begin{eqnarray}
%%%%% begin : ampl6spppms
&& -{i\sandmp5.{\ksl_1\Ksl_{234}}.5^2
     \over K_{2345}^2 \spa2.3 \spa3.4 \spa4.5
           \sandmp2.{\ksl_1\Ksl_{234}}.5}
+{i m_s^2 \sand5.{\ksl_6}.4^2 
       \sandpm4.{\Ksl_{23}\ksl_1}.2
   \over L_{12} L_{456} L_{56} \spa2.3 \spb4.5 \sandmp5.{\ksl_6\Ksl_{45}}.3}
\nonumber\\
&& - {i m_s^2 \spa2.3 \sandmp5.{\ksl_6\Ksl_{34}}.5^2 \sand5.{\Ksl_{34}}.2
   \over 
  K_{345}^2 L_{12} \spa2.3\spa3.4\spa4.5
     \sandmp5.{\ksl_6\Ksl_{345}}.2\sandmp5.{\ksl_6\Ksl_{45}}.3
  }
%%%%% end : ampl6spppms
\end{eqnarray}
in agreement with ref.~\cite{MassiveRecursion} up to an irrelevant overall 
phase\footnote{After correcting the sign 
in the second term of eqn.~(3.18) of ref.~\cite{MassiveRecursion}.}.

We have verified that \eqn{LastMinusResult} has the correct
collinear limits, and have also verified it numerically against a
set of light-cone recurrence relations up to $n=12$.

\section{The $A(1_s,2^+,\ldots,m^-,(m+1)^+,\ldots,n_s)$ Amplitude}
\label{GeneralMinusSection}

In this section, we will generalize the result from the previous section,
and obtain an expression for the amplitude with a negative-helicity
gluon not color-adjacent to one of the massive scalar legs.
Here, we will use an $(m,m-1)$ shift,
\begin{eqnarray}
&|m^-\rangle &\rightarrow |m^-\rangle - z\,|(m-1)^-\rangle \,,\nonumber\\
&|(m-1)^+\rangle &\rightarrow |(m-1)^+\rangle + z\,|m^+\rangle \,, 
\label{GeneralMinusSpinorShift}
\end{eqnarray}
Our target amplitude is then given by the following recursive form
\begin{eqnarray}
&&\hspace*{-0.8cm}A_n(1_s,2^+,\ldots,m^-,(m+1)^+,\ldots,(n-1)^+,n_s)
\nonumber\\
&=&A_m(1_s,2^+,\ldots,\widehat{(m-1)}{}^+,-\hat{K}^-_{1\cdots (m-1),s})
\frac{1}{K^2_{1\cdots (m-1)}-m_s^2}
\nonumber\\
&& \hphantom{ A_m(1_s,2^+) }\times
A_{n-m+2}(\hat{K}^+_{1\cdots (m-1),s},\hat{m}^-,(m+1)^+,\ldots,n_s)
\nonumber\\
&&+\sum_{i=m+1}^{n-1}A_{n+m-i}(1_s,2^+,\ldots ,\widehat{(m-1)}{}^+,
                  \hat{K}^+_{(i+1)\cdots (m-1)},(i+1)^+,\ldots,n_s)
\frac{1}{K^2_{(i+1)\cdots (m-1)}}
\nonumber\\
&&\hphantom{A_s(l_1^+,1^+,\ldots,m^-,\ldots,n^+,l_2^-)}\times 
           A_{i-m+2}(\hat{K}^-_{(i+1)\cdots(m-1)},\hat{m}^-,(m+1)^+,\ldots,i^+)
\nonumber\\
&&+A_3((m-2)^+,\widehat{(m-1)}{}^+,-\hat{K}^-_{(m-2)(m-1)})
\frac{1}{K^2_{(m-2)(m-1)}}
\nonumber\\
&&\hphantom{A_s(1_s,2^+)}
\times A_{n-1}(1_s,2^+,\ldots,(m-3)^+,\hat{K}^+_{(m-2)(m-1)},\hat{m}^-,\ldots,n_s)
\nonumber\\
&\equiv&U_1(1_s,2^+,\ldots,m^-,\ldots,(n-1)^+,n_s)
+A_3((m-2)^+,\widehat{(m-1)}{}^+,-\hat{K}^-_{(m-2)(m-1)})
\nonumber\\
&&\times
\frac{1}{K^2_{(m-2)(m-1)}}
  A_{n-1}(1_s,2^+,\ldots,(m-3)^+,\hat{K}^+_{(m-2)(m-1)},\hat{m}^-,\ldots,n^+_s)
\label{eq:the_definition_of_the_m_amp}
\end{eqnarray}
where as indicated $U_1$ is defined as the first two terms.
It depends only on gluonic amplitudes, 
all-plus massive-scalar amplitudes and $m=1$
lone-negative-helicity massive-scalar amplitudes. We obtained all-multiplicity
solutions for these in previous sections, and from them we can
obtain the all-$n$ form of this function.  The $U_1$ term corresponds
to the $l_1=2$ terms in \eqn{GeneralMinusResult}.

For the last term in \eqn{eq:the_definition_of_the_m_amp} 
we can, as before, iterate the recurrence relation
\eqn{eq:the_definition_of_the_m_amp}, whereupon we obtain
\begin{eqnarray}
&&\hspace*{-0.8cm}A_n(1_s,2^+,\ldots,m^-,(m+1)^+,\ldots,(n-1)^+,n_s)
\nonumber\\
&=&\sum_{j=1}^{m-2}\prod_{r=2}^{j}
\frac{iA_3\bigl(-\Kh_{[r]}^-,(m-r)^+,\dKh^+_{[r-1]}\bigr)}
{K^2_{(m-r)\cdots(m-1)}}
\nonumber\\
&&\times U_1\bigl(1_s^+,2^+,\ldots,(m-j-1)^+,\Kh^+_{[j]},\hat{m}^-,\ldots,n_s\bigr),
\label{eq:recursive_sum_form_of_m}
\end{eqnarray}
where $\Kh_{[r]}$ and $\dKh_{[r]}$ are
 as defined in \eqn{eq:shifted_K_mom_def}, and where we
have used
\begin{eqnarray}
A_{n-m+3}(1_s,\Kh_{[m-2]}^+,\hat{m}^-,\ldots,n_s)=U_1(1_s,\Kh_{[m-2]}^+,\hat{m}^-,\ldots,n_s).
\end{eqnarray}

\def\fb{\overline{\kern -0.15em f}}
\def\Gb{\overline{G}}
\def\spacer{\hphantom{\times \Bigg()}\hskip 3mm}
The amplitude we seek is now defined entirely in terms of $U_1$. Using
this iterated recurrence form of the amplitude along with the explicit
form of $U_1$ and also \eqn{eq:A_3_product_result} we arrive at our
final form for the amplitude,
\begin{eqnarray}
&&\hspace*{-0.8cm}A_n(1_s,2^+,\ldots,m^-,\ldots,(n-1)^+,n_s) =
-\frac{i}{
\spar2..{(n-1)}}
\nonumber\\
&&\times 
\sum_{l_1=2}^{m-1}\sum_{l_2=m}^{n-2}
\Bigg(
%%%%% begin : amplnmsummand
% each sandxx expression on one line for back conversion
\frac{f_0(n,m;l_1)\fb_0(n,m;l_2)}
   {\sandmp{m}.{\Ksl_{1\cdots m} \Ksl_{2\cdots(l_2+1)}+\Ksl_{2\cdots m}\ksl_{1}}.{(l_2+1)}}
\nonumber\\
&&\spacer
\times \frac{\sandmp{m}.{\Ksl_{1\cdots m}\Ksl_{(m-l_1+1)\cdots m}}.{m}^2
             \sandmp{m}.{\Ksl_{1\cdots m} \Ksl_{m\cdots (l_2+1)}}.{m}^2}
{\sandmp{m}.{\Ksl_{m\cdots (n-1)}\Ksl_{(m-l_1+1)\cdots n}+\ksl_{n} \Ksl_{(m-l_1+1)\cdots (n-1)}}.{(m-l_1+1)}}
\label{GeneralMinusResult}\\
&&\spacer\times
  \frac{1}{\sandmp{m}.{\Ksl_{m\cdots (l_2\!+\!1)}\bigl(\Ksl_{(l_2\!+\!2)\cdots (n\!-\!1)}\Ksl_{(m\!-\!l_1\!+\!1)\cdots n}+\ksl_{n} \Ksl_{(m\!-\!l_1\!+\!1)\cdots (n\!-\!1)}\bigr)\Ksl_{(m\!-\!l_1\!+\!1)\cdots m}}.{m}}
\nonumber\\
&&\spacer\hskip-4mm
 +\frac{f_1(n,m,l_1,l_2+1;l_1)f_2(n,m,l_1,l_2;l_2+1)
   \sandmp{m}.{\Ksl_{m\cdots (l_2+1)}\Ksl_{(m-l_1+1)\cdots m}}.{m}^4}
{K_{(m-l_1+1)\cdots (l_2+1)}^2
 \sandmp{(l_2+1)}.{\Ksl_{(m-l_1+1)\cdots (l_2+1)}\Ksl_{(m-l_1+1)\cdots m}}.{m}}
\nonumber\\
&&\spacer\times
\frac{1}{\sandmp{m}.{\Ksl_{m\cdots (l_2+1)}\Ksl_{(m-l_1+1)\cdots (l_2+1)}}.{(m-l_1+1)}}
%%%%% end : amplnmsummand
\Bigg),
\nonumber
\end{eqnarray}
which is valid for $2<m<n-1$.  (For $m=n-1$, one should use either 
\eqn{LastMinusResult} or~(\ref{LastMinusResult2}); for $m=2$, one should
use those formul\ae{} after reflection.)
In the above equation,
\begin{eqnarray}
f_0(n,m;i)&=&\left\{\begin{array}{ll} -1, &i=m-1;
\\
 \displaystyle
%%%%% : f0case2
  \frac{\spa{(m-i)}.{(m-i+1)} G(2,m-i+1;m)/L_{(m-i+1)\cdots n}}{L_{12}
     \sandmp{(m-i)}.{\Ksl_{(m-i+1)\cdots (n-1)}\Ksl_{m\cdots n}+\ksl_n \Ksl_{m\cdots (n-1)}}.{m}},
%%%%% : f0case2
&2\leq i < m-1.
\end{array}\right.
\nonumber\\
\\
\fb_0(n,m;i)&=&\left\{\begin{array}{ll} 1, &i=n-2;
\\
 \displaystyle{\spa{(i+1)}.{(i+2)}
 \Gb(n;i+1,n-1;m) \over L_{1\cdots (i+1)}
  \sandmp{(i+2)}.{\Ksl_{2\cdots (i+1)}\Ksl_{1\cdots m}
                  +\ksl_1 \Ksl_{2\cdots m}}.{m}L_{n-1,n}},
& m\leq i < n-2.
\end{array}\right.\nonumber
\end{eqnarray}
and
\begin{eqnarray}
f_1(n,m,l_1,l_2;i)&=&\left\{\begin{array}{ll}
\displaystyle
\frac{G_0(2,m-l_1+1;l_2,n-1;2;m)}{\Bmp{m^-|\s
    K_{m\cdots l_2}\s K_{2\cdots l_2}\s k_1 \s K_{2\cdots m}|m^+}},
&i=m-1;
\\
\displaystyle \vphantom{\sum_a^b}
{\spa{(m-i)}.{(m-i+1)} G_1(2,m-i+1;l_2,n-1;2;m)\over L_{12}\Bmp{(m-i)^-|\s
  K_{(m-i+1)\cdots l_2}\s K_{m\cdots l_2}|m^+}},
&2\leq i < m-1.
\end{array}\right.
\nonumber\\
\\
f_2(n,m,l_1,l_2;i)&=&\left\{\begin{array}{ll}
\displaystyle
\frac{-1}{\Bmp{m^-|\s K_{m\cdots (n-1)}\s k_n\s K_{(m-l_1+1)\cdots (n-1)}\s
K_{(m-l_1+1)\cdots m}|m^+}},
&i=n-1;
\\
\displaystyle
{-\spa{i}.{(i+1)}
 \over L_{(n-1)n}\Bmp{(i+1)^-|\s K_{(m-l_1+1)\cdots i} \s K_{(m-l_1+1)\cdots m}|m^+}},
&m+1\leq i < n-1.
\end{array}\right.\nonumber
\end{eqnarray}
The attentive reader will notice that the dimensionality of the
expressions for $f_1$ and $f_2$ is different in each of the two cases.
Nonetheless, all terms in \eqn{GeneralMinusResult} have the same dimension,
because this difference is compensated by the inapplicability of the
first replacement in \eqn{eq:replacment_ids_2} when $l_2=n-2$.

\def\pred{\mathop{\rm pred}\nolimits}
\def\succ{\mathop{\rm succ}\nolimits}
We have also defined
\begin{eqnarray}
&&\hspace*{-0.8cm}G_u(a_1,b_1;a_2,b_2;c;m) =
\nonumber\\
&&\sum^{\lfloor (b_1+b_2-(a_1+a_2)+c)/2 \rfloor}_{j=1}
     \mskip -30mu (-m_s^2)^j
  \mathop{{\sum}'}_{\{w_i\}_{i=1}^{j-1}\atop w_i\in S_i}
\mskip -10mu
\biggl(\prod_{r=1}^{j-1}\frac{
\sand{w_r}.{\Ksl_{(w_r+1)\cdots (w_{r+1}-1)}}.{w_{r+1}}}{L_{1\cdots (w_r-1)}L_{1\cdots w_r}}\biggr)
\\
&&\hphantom{ \sum^{\lfloor (b_1+b_2-(a_1+a_2)+n)/2 \rfloor}_{j=1}(-m_s^2)^j
  \sum_{\{w_k\}_{k=1}^{j-1}\atop w_i=w_{i-1}+2}^{b_2-2} } \times 
\left\{\begin{array}{ll}
\displaystyle
 \sandpp{w_1}.{\Ksl_{a_1\cdots (w_1-1)}\ksl_1\Ksl_{2\cdots m}}.{m}, &u=0,
\\
\displaystyle
 \sandpm{w_1}.{\Ksl_{a_1\cdots (w_1-1)}\ksl_1}.{a_1}, & u=1,
\end{array}\right\}     \Bigg|_{w_0=a_1-1\atop w_j=b_2},\nonumber
\end{eqnarray}
where the sum for each $w_i$ is over a set of momenta $S_i$, defined
as follows.  Define the set $S_0=\{a_1,\ldots,b_1,a_2,\ldots,b_2\}$,
let $\succ(a)$ be the element following $a$, and $\pred(a)$ the element
preceeding $a$ in $S_0$.
The set $S_1$ is then defined by omitting the first and
last two elements of $S_0$,
\begin{equation}
S_1 = \{\succ(a_1),\ldots, b_1,a_2,\ldots,\pred(\pred(b_2))\}.
\end{equation}
Each subsequent set $S_i$ depends on the value of $w_{i-1}$; it will
contain all elements in $S_0=\{a_1,\ldots,b_1,a_2,\ldots,b_2\}$ following
the element after $w_{i-1}$, and through two elements prior to $b_2$.
  This is the generalization of the sum
for $w_i$ starting at $w_{i-1}+2$ (and ending at $w_{i+1}-2$).  That is, 
\begin{equation}
S_i = \{\succ(\succ(w_{i-1})),\ldots,\pred(\pred(b_2))\}.
\end{equation}
Note, however, that
the sums of momenta $K_{i\cdots j}$ are over 
consecutive momenta, {\it not\/} restricted to $S_0$.

The prime signifies that in addition, 
we must make the following
replacements whenever $w_k$ is inside a bra or a ket. 
For $w_k=b_1$,
\begin{eqnarray}
\frac{\langle w_k^-|\s K_{(w_k+1)\cdots (w_{k+1}-1)}}{L_{1\cdots w_k}}
&\rightarrow&
\frac{\langle m^-|\s K_{m\cdots a_2}\s K_{b_1\cdots a_2}
      \s K_{w_k\cdots (w_{k+1}-1)}}{\Bmp{m^-|\s K_{b_1\cdots m}\left(\s
    K_{1\cdots (b_1-1)}\s K_{2\cdots a_2} + \s K_{2\cdots (b_1-1)}\s k_1\right) \s K_{m\cdots a_2}|m^+}}
\nonumber\\
\s
K_{(w_{k-1}+1)\cdots (w_{k}-1)}|w_{k}^-\rangle&\rightarrow&
 -\s {K}_{(w_{k-1}+1)\cdots (w_{k}-1)}\s K_{b_1\cdots m}|m^+\rangle
\label{eq:replacment_ids_1}
\end{eqnarray}
and for $w_k=a_2$,
\begin{eqnarray}
\frac{\langle w_{k}^-|\s
  K_{(w_k+1)\cdots (w_{k+1}-1)}}{L_{1\cdots (w_k-1)}}&\rightarrow&
\frac{\langle m^-|\s K_{b_1\cdots m}\s K_{b_1\cdots a_2}\s K_{(w_k+1)\cdots (w_{k+1}-1)}}{\Bmp{m^-|\s K_{b_1\cdots m}\left(\s
    K_{1\cdots (b_1-1)}\s K_{2\cdots a_2} + \s K_{2\cdots (b_1-1)}\s k_1\right) \s K_{m\cdots a_2}|m^+}}
\nonumber\\
\s K_{(w_{k-1}+1)\cdots (w_{k}-1)}|w_{k}^-\rangle&\rightarrow& \s
K_{(w_{k-1}+1)\cdots w_{k}}\s K_{m\cdots a_2}|m^+\rangle
\label{eq:replacment_ids_2}
\end{eqnarray}
These replacements should also be applied to $\langle w_1^+|$ 
in the last factor. That is if $w_1 = b_1$,
\begin{eqnarray}
\langle w_{1}^+ |\s
K_{a_1\cdots (w_{1}-1)}&\rightarrow&
\langle m^-|\s K_{b_1\cdots m}\s {K}_{a_1\cdots (w_{1}-1)},
\label{eq:replacment_ids_1a}
\end{eqnarray}
and if $w_1 = a_2$,
\begin{eqnarray}
\langle w_{1}^+|\s K_{a_1\cdots (w_{1}-1)}
&\rightarrow& -\langle m^-|
\s K_{m\cdots a_2}\s K_{a_1\cdots w_{1}}.
\label{eq:replacment_ids_2a}
\end{eqnarray}

This definition of $G$ is a generalization of that given in
\eqn{eq:def_simple_G}, the two are related via $G(a_1,b_1;m)\equiv
G_1(a_1,b_1;b_1,b_1;1;m)$ as
$\{a_1,\ldots,b_1\}\bigcup\{b_1\}=\{a_1,\ldots,b_1\}$ and only the second
replacement from \eqn{eq:replacment_ids_1} is relevant.  Similarly
$\Gb$ is defined as
\begin{eqnarray}
\Gb(n;a_2,b_2;m)&=&\sum_{j=1}^{\lfloor (b_2-a_2+1)/2 \rfloor }(-m_s^2)^j
\sum_{\{w_i\}^{j-1}_{i=1}=a_2+2\atop w_{i}=w_{i+1}+2}^{b_2-1}
\biggl(\prod_{r=1}^{j-1}\frac{\Bmp{w_r^-|\s K_{(w_{r+1}+1)\cdots (w_{r}-1)}|w_{r+1}^-}}
{L_{w_r\cdots n}L_{(w_{r}+1)\cdots n}}\biggr)
\\
&&\hphantom{ \sum_{j=1}^{\lfloor (b_2-a_2+1)/2 \rfloor }(-m_s^2)^j
\sum_{\{w_k\}^{j-1}_{k=1}=a_2+2}^{b_2-1} }\times
  \sandpm{w_1}.{\Ksl_{(w_1+1)\cdots (n-1)}\ksl_{n}}.{b_2}
    \Bigg|_{\langle w_j^+|=\langle m^-|\Ksls_{m\cdots a_2}\atop w_j=a_2}.
\nonumber
\end{eqnarray}
%where the sum for each $w_k$ is over the sequence $\{a_2,\ldots,b_2\}$.

There is only a single massless contribution in \eqn{GeneralMinusResult};
it arises from the $(l_1=m-1,l_2=n-2)$ term and is given by
\begin{eqnarray}
%%%%% begin : amplnmassless
&&  \frac{i \sandmp{m}.{\ksl_1 \Ksl_{2\cdots m}}.{m}^2}
       {\spar2..{(n-1)}
       \sandmp{m}.{\Ksl_{m\cdots (n-1)}\ksl_{n}}.{(n-1)}
     }
\nonumber\\
&& \hskip 5mm\times
  \frac{\sandmp{m}.{\ksl_n \Ksl_{m\cdots (n-1)}}.{m}^2}
  {\sandmp{m}.{\Ksl_{m\cdots (n-1)}\ksl_{n}\Ksl_{2\cdots (n-1)}\Ksl_{2\cdots m}}.{m}
      \sandmp{m}.{\Ksl_{2\cdots m}\ksl_{1}}.{2}}
%%%%% end : amplnmassless
\end{eqnarray}
which is readily seen to reduce to the expected MHV amplitude in the massless
limit. Using \eqn{GeneralMinusResult} we can derive the form of the amplitude
in the five point, $m=3$ case
\begin{eqnarray}
&&\hspace*{-0.8cm} A_5(1_s,2^+,3^-,4^+,5_s)
\nonumber\\
&=&
%%%%% begin : ampl5spmps
-\frac{i\sand3.{\ksl_1}.2^2\sand3.{\ksl_5}.4^2}{L_{12}
\spa2.3\spa3.4 L_{45}\sandpm4.{\ksl_5\Ksl_{34}}.2}
+\frac{i m_s^2 \spb2.4^4}{K_{2\cdots 4}^2 \spb2.3\spb3.4
  \sandpm4.{\ksl_5\Ksl_{34}}.2},
%%%%% end : ampl5spmps
\end{eqnarray}
this result matches exactly that given in ref.~\cite{MassiveRecursion}.
For the six point case with $m=4$ we have
\def\spacer{\hskip 10mm}
\begin{eqnarray}
&&A_s(1_s,2^+,3^+,4^-,5^+,6_s)
=\cr
%%%%% begin : ampl6sppmps
&&\spacer\frac{i\sandmp4.{\ksl_1\Ksl_{23}}.4^2 \sand4.{\ksl_6}.5^2}
{\spa2.3\spa3.4\spa4.5 L_{56}\sandpp5.{\ksl_6\Ksl_{2\cdots 5}\Ksl_{23}}.4
 \sandmp4.{\Ksl_{23}\ksl_1}.2}
\nonumber\\
&&\spacer-\frac{i m_s^2 \spb2.3\sand4.{\Ksl_{56}}.3^2\sand4.{\ksl_6}.5^2}
{L_{12}L_{4\cdots 6}L_{56}\spa4.5 \sandmp2.{\ksl_1\Ksl_{23}}.4\sandpm5.{\ksl_6\Ksl_{45}}.3}
\nonumber\\
&&\spacer-\frac{i m_s^2\spb3.5^4\sandpm5.{\Ksl_{2\cdots 4}\ksl_1}.2}
{L_{12}\spb3.4\spb4.5\sand2.{\Ksl_{34}}.5\sandpm5.{\ksl_6\Ksl_{45}}.3
 K_{3\cdots 5}^2}
\nonumber\\
&&\spacer -\frac{i m_s^2\sand4.{\Ksl_{23}}.5^4}
{K_{2\cdots 5}^2 K_{2\cdots 4}^2\spa2.3\spa3.4
\sandpp5.{\ksl_6\Ksl_{2\cdots 5}\Ksl_{23}}.4
\sand2.{\Ksl_{34}}.5}.
%%%%% end : ampl6sppmps
\label{eq:n=4_m=3_test_case}
\end{eqnarray}
Although this result has a different form from that given 
in ref.~\cite{MassiveRecursion}, we have checked that the two
agree numerically up to an overall 
phase\footnote{After correcting a typographical error 
in eqn.~(3.21) of ref.~\cite{MassiveRecursion}.}.
We have also compared the result~\eqn{GeneralMinusResult} numerically
to amplitudes computed via light-cone recurrence relations, and found complete
agreement through $n=12$.

\section{Conclusions}

On-shell recursion relations have emerged as a powerful method for deriving
analytic expressions for tree-level amplitudes.  In this paper, we have
given all-multiplicity solutions
to the BGKS recursion relation for amplitudes with a massive scalar
pair for the two simplest helicity configurations.  These are the
amplitudes with an arbitrary number of positive-helicity gluons, and
either no or one gluon of negative helicity.  (Of course, the corresponding
amplitudes with an arbitrary number of negative-helicity gluons, and up
to one gluon of positive helicity, can be obtained by spinor conjugation.)
We have checked the principal results, eqs.~(\ref{AllPlusResult}), 
(\ref{LastMinusResult}), and~(\ref{GeneralMinusResult}), against
a numerical implementation of the older light-cone recursion relations,
through $n=12$.

The primary application of these results is to computations of one-loop
QCD amplitudes within the unitarity-based method.  
The massive
scalar amplitudes are equivalent to amplitudes with the scalar legs
computed fully in $D$ dimensions.  These in turn can be used to obtain
rational terms in amplitudes, which have no cuts in four dimensions,
but do have cuts in $D$ dimensions.  
They should also be
useful for proving general factorization properties in complex momenta
upon which the recursion-relations part of the unitarity-bootstrap combined
approach relies~\cite{OneLoopRationalRecursion}.
For such applications, a relatively
compact analytic form is desirable, and the recursion relations allow
one to obtain such a form.

While the all-plus amplitude did require guesswork for an all-$n$ ansatz,
it is interesting to note that the amplitudes with a negative-helicity
gluon could then be derived without having to guess an ansatz
for the general form.  This would not be the case using the older,
off-shell recursion relations~\cite{Recursion,LightConeRecursion}.  
While the over-all complexity of the amplitudes would increase
with increasing number of negative-helicity gluons, the same approach
of unwinding three-point vertices could be used in such cases as well.
These too would not require new ans\"atze for massive-scalar amplitudes.
The calculations in this paper
also illustrate the extent to which calculations beyond those in 
purely massless gauge theories are feasible.  We would expect, for
example, that amplitudes with massive quarks or vector bosons instead
of massive scalars would yield to a calculation of similar nature.

%%%%%%%%%%%%%%%%%%%%%%%%%%%%

\section*{Acknowledgments}

We thank Academic Technology Services at UCLA for computer support,
and Z.~Bern and L.~Dixon for helpful comments.

\appendix

\section{Computational Complexity of Berends--Giele Recursion Relations}
\label{ComputationalComplexity}

In order to use QCD amplitudes in evaluating experimental data,
one must evaluate them numerically.  How many operations are required
to do so?
Even at tree level, each amplitude, and indeed
each different helicity amplitude, require 
a factorial number of Feynman diagrams.  This might seem to imply
that a factorial number of operations are required; as we shall
show here, a better organization of the calculation can dramatically
reduce this complexity.

The complete squared amplitude can be written in terms of
color-ordered amplitudes.  At leading order in the number of
colors, which is all we will consider here, computing the
amplitude does require a sum over a factorial number of 
color orderings.
However, for integrated quantities, one can use the symmetry of
final-particle phase space to reduce this to a sum over a 
linear number of orderings.  We are thus left to consider the
cost of computing a given color-ordered amplitude.

There are $\Ord(2^{n/2})$ independent helicity amplitudes, so 
clearly the complexity of computing all of them must scale
at least exponentially in the number of external legs.  
There are an exponential number
of color-ordered Feynman diagrams for any given helicity
amplitude, which would suggest that we still require
an exponential number of operations.

In this appendix, we show that the number of operations 
required to evaluate a typical helicity amplitude is only
polynomial in the number of external legs, so long as the
calculation is organized properly.  To do so, we will
make use of a light-cone version~\cite{LightConeRecursion,CSWCurrent}
of the Berends--Giele recursion relations~\cite{Recursion}.  
(The complexity
is polynomial for any version of this type of off-shell recursion
relation, suitably organized; only the prefactors differ.)
In our estimates, we take each basic operation --- addition,
multiplication, division --- to have constant complexity,
that is $\Ord(1)$.

Let $J^{\pm}(1,\ldots,n)$ denote the (amputated) 
current with $n$ on-shell
external legs, and one off-shell external leg of the given helicity
(defined using a light-cone gauge vector $q$).  We can write
a recursion relation for this current,
\begin{equation}
\begin{array}{rl}
\displaystyle J^{\sigma}&(1,\ldots,n) =\\
& \displaystyle 
-\sum_{j_1=1}^{n-1} \sum_{\sigma_{1,2}=\pm}
{V_3^{\sigma\sigma_1\sigma_2}
 \over K_{1\ldots j_1}^2 K_{(j_1+1)\ldots n}^2}
\, J^{-\sigma_1}(1,\ldots,j_1) J^{-\sigma_2}(j_1\!+\!1,\ldots,n)
\\
&\displaystyle 
+\sum_{j_1=1}^{n-2} \sum_{j_2=j_1+1}^{n-1}
\sum_{\sigma_{1,2,3}=\pm}
{i V_4^{\sigma\sigma_1\sigma_2\sigma_3} 
\over K_{1\ldots j_1}^2 K_{(j_1+1)\ldots j_2}^2 K_{(j_2+1)\ldots n}^2}
\,J^{-\sigma_1}(1,\ldots,j_1)
\\
& \displaystyle \hphantom{ -{i\over K_{1,n}^2} \Biggl[]}
  \vphantom{ \sum_{j=1}^{n-2} }
 \hskip 10mm\times J^{-\sigma_2}(j_1\!+\!1,\ldots,j_2)
 J^{-\sigma_3}(j_2\!+\!1,\ldots,n)
\end{array}
\end{equation}
In order to compute an $n$-point amplitude, $A_n$, we compute
$J^\sigma(1,\ldots,n-1)$ with the $n$-th leg on-shell instead of off-shell.
This in turn requires the computation of currents with fewer
on-shell external legs.
We can organize this computation as follows: first compute all
the required three-point currents; then the four-point currents,
and so on through the desired $n$-point current.  At each
stage, only previously-computed
lower-point currents appear, so we need compute only the vertices
and propagators, then multiply factors and perform the sums.  
In practice, it is probably more convenient to use caching
rather than precomputation.  That is, use the recursive formula, but during
the recursive descent, record the (numerical) value of
each newly-computed current.  The next time that current is required,
use the previously-computed value instead of computing it anew.

For each
$j$-point current contributing to the final
$n$-point current, the computation is clearly dominated by 
the four-point vertices and associated double sums.  
We can also precompute all momentum sums that appear in our
computation.  These are always sums of consecutive momenta,
so there are $\Ord(n^2)$ different ones, and this precomputation
is of $\Ord(n^3)$ in operations.  We can also precompute
required Lorentz and spinor products, which requires
$\Ord(n^2)$ operations.
Each four-point vertex then takes a constant number of operations
to compute.  There are $\Ord(j^2)$ different terms, each requiring
a constant number of operations, and so the computation of
each $j$-point current requires $\Ord(j^2)$ additional operations.

For a given $j$, we need $n-j+1$ different
$j$-point currents: $J(1,\ldots,j-1)$
through $J(n-j+1,n-1)$.  The overall computational complexity
is thus of order,
\begin{equation}
\sum_{j=3}^{n} (n-j+1) j^2
\sim \Ord(n^4),
\end{equation}
which is indeed polynomial.

This is possible because we can re-use information: 
the computation of $J_{17}$, for example, requires knowledge of
both $J_{10}(1,\ldots,10)$ and $J_5(1,\ldots,5)$.  The computation
of $J_{10}(1,\ldots,10)$ in turn requires knowledge of 
$J_5(1,\ldots,5)$, but this five-point current is the {\it same\/}
as the one appearing in the top-level sum for $J_{17}$, and thus
need be computed only once.  This saving would not be possible
if we insisted on writing out an analytic expression for $J_{17}$
in terms of spinor products, because the
expression for $J_5$ would then appear
(and effectively be computed) many times.

We can obtain a more precise count of the number of operations 
as follows.  Denote the number of operations required to compute
a three-vertex $V_3$ by $c_3$, and that for a four-vertex $V_4$
by $c_4$.  If all helicity arguments to a three-vertex are identical,
it vanishes, so for a given choice of the current's helicity
$\sigma$,
there are only three different helicity choices we must compute.
A straightforward organization of the sums
requires four multiplications or divisions and one
addition for each one, for example,
\begin{equation}
{V^{-+-} J_1^{-} J_2^{+}\over K_1^2 K_2^2}
+{V^{--+} J_1^{+} J_2^{-}\over K_1^2 K_2^2}
+{V^{-++} J_1^{-} J_2^{-}\over K_1^2 K_2^2},
\end{equation}
where $J_1 \equiv J(1,\ldots,j_1)$, $K_1\equiv K_{1\ldots j_1}$, 
etc.
It is possible to reduce this slightly by combining terms as follows,
\begin{equation}
{1\over K_1^2 K_2^2} \bigl[V^{-+-} J_1^{-} J_2^{+}
+J_2^{-} (V^{--+} J_1^{+} +V^{-++} J_1^{-})\bigr],
\end{equation}
giving instead a count of $3 c_3+10$ for a typical term,
or $(3 c_3+10)(j-2)$ for a $j$-point current.
(This is still an overestimate of the optimal
count, because we ignore the reduction due to terms containing
a two-point current as it depends on the choice of external
helicities.)

Only four-vertices with two negative-helicity and
two positive-helicity arguments are non-vanishing, so here we again
have three different helicity configurations for each choice of
the current's helicity. This would give six multiplications
and one addition for each term in the double sum over $j_1$
and $j_2$ and $3 (c_4+7) (j-2)(j-3)/2$ overall.  Here,
we can reorganize not only each term but the double sum 
as well,
\begin{equation}
\begin{array}{rl}
&\displaystyle\sum_{j_1=1}^{j-3} \sum_{j_2=j_1+1}^{j-2} 
\Bigl[
{V^{-++-} J_1^{-} J_2^{-} J_3^{+}\over K_1^2 K_2^2 K_3^2}
+{V^{-+-+} J_1^{-} J_2^{+} J_3^{-}\over K_1^2 K_2^2 K_3^2}
+{V^{--++} J_1^{+} J_2^{-} J_3^{-}\over K_1^2 K_2^2 K_3^2}
\Bigr] =
\cr &\displaystyle\hphantom{=}\sum_{j_1=1}^{j-3} {1\over K_1^2} 
\biggl\{ J_1^{-}\mskip -2mu
  \sum_{j_2=j_1+1}^{j-2}
     {1\over K_2^2 K_3^2} \bigl[
              V^{-++-} J_2^{-} J_3^{+}
              +V^{-+-+} J_2^{+} J_3^{-}
              \bigr]
%\cr & \hphantom{ =\sum_{j_1=1}^{j-3} {1\over K_1^2}\biggl\{ }
+ J_1^{+} \mskip -2mu\sum_{j_2=j_1+1}^{j-2}
{V^{--++} J_2^{-} J_3^{-}\over K_2^2 K_3^2}
\biggr\},
\end{array}
\end{equation}
giving an overall operation count of 
\begin{equation}
{3\over2}(c_4+4) (j-2)(j-3) + 4 (j-3)
\end{equation}
(again ignoring the reduction due to two-point currents).

Overall, we then obtain an operation count of
\begin{equation}
\begin{array}{rl}
&\displaystyle\sum_{j=3}^{n} (n-j+1)\Bigl[
  (3 c_3+10)(j-2)
+{3\over2}(c_4+4) (j-2)(j-3) + 4 (j-3)
\Bigr] 
\cr &\displaystyle= 
{n(n-1)(n-2)(n-3)\over 8}  c_4
+{n(n-1)(n-2)\over2}  c_3
+{(n-1)(n-2)(n+3)(3n-4)\over6} 
\end{array}
\end{equation}

In contrast to the off-shell recursion relations analyzed here,
in on-shell recursion relations, the reference momenta will change
as the recursion descends.  It thus seems likely that the
computational complexity will still be exponential for a typical
color-ordered helicity amplitude.  Of course, for special
helicity configurations or for a moderate number of external
legs, the new representations may provide a more efficient
computational method.

%%%%%%%%%%%%%%%%%%%%%%%%%%%%%%%%%%%%%%%%%%%

\end{document}